# Estimation of Spatial-Temporal Gait Parameters based on the Fusion of Inertial and Film-Pressure Signals

Cheng Wang[1,2,3,4], Xiangdong Wang[1,3], Zhou Long[4], Tian Tian[5], Mingming Gao[6], Xiaoping Yun[6], Yueliang Qian[1,3], and Jintao Li[1,3]
[1]Institute of Computing Technology(ICT), Chinese Academy of Sciences(CAS), Beijing, China
[2]University of Chinese Academy of Sciences(UCAS), Beijing, China
[3]Beijing Key Laboratory of Mobile Computing and Pervasive Device, Beijing, China
[4]Luoyang Institute of Information Technology Industries, Luoyang, China
[5]Beijing Chao-Yang Hospital, Capital Medical University, Beijing, China
[6]China Rehabilitation Research Center, Beijing, China
{wangcheng01, xdwang, longzhou}@ict.ac.cn, tt791201@163.com, wishyouokgmm@163.com, yunxiaoping@hotmail.com, {ylqian, jtli}@ict.ac.cn

*Abstract*—Gait analysis (GA) has been widely used in physical activity monitoring and clinical contexts, and the estimation of the spatial-temporal gait parameters is of primary importance for GA. With the quick development of smart tiny sensors, GA methods based on wearable devices have become more popular recently. However, most existing wearable GA methods focus on data analysis from inertial sensors. In this paper, we firstly present a two-foot-worn in-shoe system (Gaitboter) based on low-cost, wearable and multimodal sensors' fusion for GA, comprising an inertial sensor and eight film-pressure sensors with each foot for gait raw data collection while walking. Secondly, a GA algorithm for estimating the spatial-temporal parameters of gait is proposed. The algorithm fully uses the fusion of two kinds of sensors' signals: inertial sensor and film-pressure sensor, in order to estimate the spatial-temporal gait parameters, such as stance phase, swing phase, double stance phase, cadence, step time, stride time, stride length, velocity. Finally, to verify the effectiveness of this system and algorithm of the paper, an experiment is performed with 27 stoke patients from local hospital that the spatial-temporal gait parameters obtained with the system and the algorithm are compared with a GA tool used in medical laboratories. And the experimental results show that it achieves very good consistency and correlation between the proposed system and the compared GA tool.

*Keywords—gait analysis, gait parameters, inertial sensor, film-pressure sensor, wearable device*

## I. Introduction

Walking is one of the most common human physical activities. Gait analysis (GA) is the systematic research of human walking motion, and it is playing significant role in health diagnostics[1] and rehabilitation[2] for tasks such as evaluating balance and mobility in abnormal gait patients before treatment and monitoring recovery status after treatment. Gait spatial-temporal parameters are the frequently used parameters of GA and is very helpful for GA[3]. Lots of devices and approaches have been developed to assist in the study of GA. Early GA approaches, such as the timed "up & go" test-TUG[4], the Berg balance scale-BBS[5], the performance oriented mobility assessment-POMA[6] are standard test process which judges the patients' gait status through experts' observation. These GA approaches are subjective, manual, qualitative and need trained-staff involved deeply. At present, most big hospitals or rehabilitation centers are using a non-subjective, automatic, quantitative and accurate approach for GA such as multi-camera motion capture system and force platforms[7,8], but it has some disadvantages such as expensive-cost, large-occupied-area, specialized setups, operating-time-consuming, indoor-laboratory-environment, etc. Therefore it cannot meet the needs of daily or real-time monitoring for abnormal gait patients (stroke, multiple sclerosis, Parkinson, etc.)' recovery at home after leaving the hospital or rehabilitation center. Therefore, quantitative, low-cost, portable and wearable GA approaches and instruments based on tiny sensors have been paid more attention and developed quickly[9,10]. The author's previous contribution is that microphone sensors are used for wearable GA because they are tiny, low-cost, portable, etc., and a microphone sensor with a two-foot-ankle worn system can get gait information directly from footstep sounds generated by the impact between someone's foot and the floor as he or she moves around, so it achieved good results for gait temporal parameters measurements[11]. However the signals from the microphone sensors on the ankle showed some noises and it will affect the results more or less. In addition, the spatial parameters of gait are hard to be measured with only this single sensor. Film-pressure sensor based methods(foot switches, pressure insole) are used for gait temporal parameter measurement[13-15] because it's thin. By using only this sensor, the spatial parameters of gait are hard to be measured and the GA results generally are unsatisfactory for abnormal walking [16]. Most wearable GA approaches only use inertial sensor based methods[17-19] thanks to their tiny size, low-cost, not limited by the environment, and low-power-consumption. In theory, it can measure gait spatial-temporal parameters. Gait temporal parameters are mainly calculated by threshold-based peak detection methods, so it is very hard to get high detection accuracy because it's almost impossible to find a fixed threshold adapting to different kinds of conditions. Gait spatial parameters are calculated by integral computation of raw inertial signals. Namely, double integration of measured accelerations signals is needed to get the displacement or position information. Unfortunately, it is difficult to obtain accurate motion accelerations because of the presence of inertial sensor bias and measurement noise, which leads to the exponential increase of displacement errors over integration time [20,21]. Generally, this kind of errors are corrected and revised by a zero velocity update (ZVU). Current detection methods of zero velocity are mainly determined by gait temporal parameters [22,23]. Hence the accuracy of spatial parameters measurement is hard to get high.

In summary, regarding to wearable GA, microphone sensors based method and film-pressure sensor based method can't measure spatial parameters. And inertial sensor based GA method is very appropriate for the wearables and can measure gait spatial and temporal parameters in theory, but accurate measurements of gait temporal parameters is a great challenge. In our previous work, we have already tried the fusion of microphone sensor and inertial sensor for GA that the result is very good for footstep detection[12]. And here we have already known that film-pressure sensor based method can measure the temporal parameters of gait. Therefore, multimodal sensor fusion such as inertial sensor and film-pressure sensor can be a promising technology for wearable GA studies. Although several in-sole or in-shoe systems, which used the inertial sensors and film-pressure sensor, have been reported in the literature[24,25], but the gait related parameters were calculated by using the signals of each sensor respectively. Namely, there is no actual fusion when estimating the gait related parameters.

The main work in this paper is organized as follows:

Firstly, a novel two-foot-worn in-shoe system (Gaitboter) is designed and developed, which is a low-cost, wireless, wearable and multimodal gait system using inertial sensor and film-pressure sensor for gait raw data collection for each foot while walking.

Secondly, based on this system, a gait analysis algorithm is proposed for estimating the spatial and temporal parameters of gait. The algorithm fully uses the fusion of two kinds of sensors' signals: inertial sensor and film-pressure sensor.

Finally, experiments are conducted to assess the effectiveness of the proposed system and algorithm with 27 stoke patients from local hospital.

## II. SYSTEM OVERVIEW

This section mainly explains the system (Gaitboter) which includes a prototype device for gait raw data collection and a tablet computer for the handling of collected gait raw data. The system is shown in Fig. 1.

A two-foot-worn in-shoe prototype device is presented to collect inertial data and film-pressure data from each feet as it is shown in Fig.1. One 6-axis inertial sensors (3-axis accelerometer, 3-axis gyroscope) and eight film-pressure sensors are the data collection sensors for each foot/shoe.

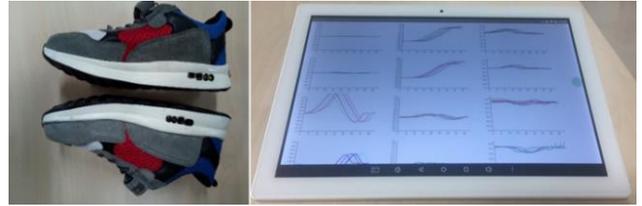

Fig. 1. Picture of the system(prototype device + tablet computer)

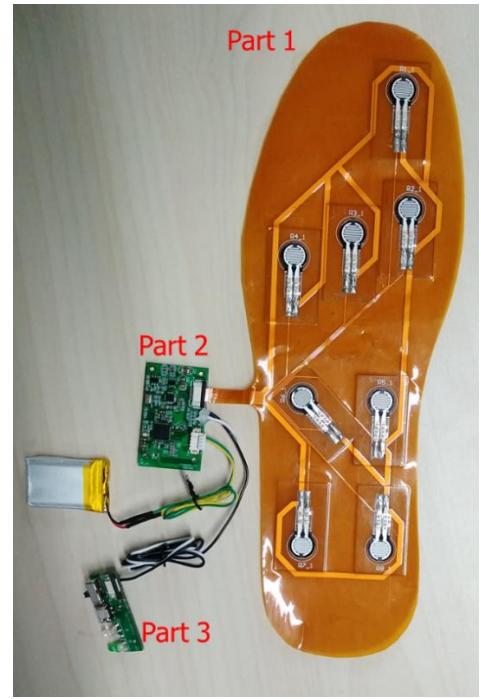

Fig. 2. Diagram of the device circuit structure

As it is shown in Fig.2 that the prototype device from the point of circuit includes three parts: part 1, part 2 and part 3. Part 1 is insole which used eight film-pressure sensors (RP-C-10, Film Sensor Technology Co., LTD. it's shown in Fig.3-a) to measure sole pressure.

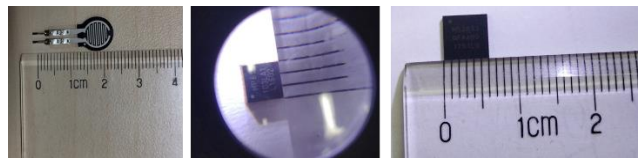

Fig. 3. Photograph of Sensors & MCU, (a) Film-pressure sensor, (b) 6-axis inertial sensor, (c)MCU

Film-pressure sensors arranged on the insole as shown in Fig.4. Numbers marked on the insole mean channel number of film-pressure sensors. Channel 1 to 4 corresponds to the forefoot(channel 1 corresponds to foot thumb), channel 5 and 6 correspond to the midfoot and channel 7 and 8 correspond to the hind foot. Part 2 is main control circuit which is arranged at the midfoot of sole(inside) as it is shown in Fig.5, and it includes one 6-axis inertial sensor (MPU6500, InvenSense Inc. it's shown in Fig.3-b) to measure acceleration and angular velocity for each foot while walking, a micro controller unit(MCU) integrated with BLE(Bluetooth Low Energy) communication (nRF52832, Nordic Semiconductor Inc. it's shown in Fig.3-c) to get the voltage value that changed by film-pressure sensors and the value of 6-axis inertial sensor, and a charging IC for charging control. Part 3 is for human interaction which is positioned at the midfoot of sole(outside). It contains LED for the indicator of device working status, switch for the power on/off of device, and USB port for charging. In addition, the protype device has the serial interface such as UART functions to communicate PC and micro-controller-unit for the program downloading and upgrading of device. The collection data by the prototype device are sent to the tablet computer with wireless communication by using Bluetooth module(BLE). And the handling method for the collected raw data is implemented at the tablet computer which will be specified in next section. The sampling rate of data collection with this device is 66 Hz. Finally, part 1, part 2, and part 3 all are embedded into sole(in-shoe). Fig.6 shows the working schematic of the presented prototype device. The connection between film-pressure sensors and MCU is ADC(Analog-to-Digital Converter), and the connection between 6-axis inertial sensor and MCU is I2C(Inter-Integrated Circuit). The prototype device's specification list is given in Table I.

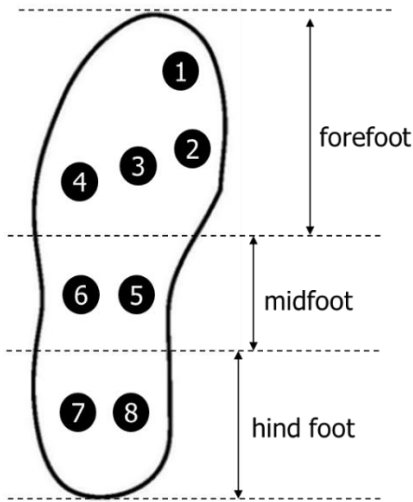

Fig. 4. Photograph of arrangement of film-sensors

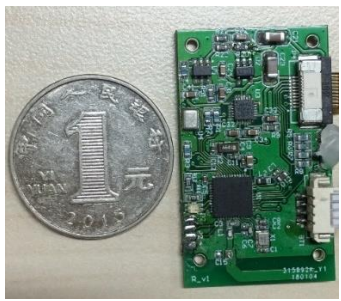

Fig. 5. Diagram of main control circuit

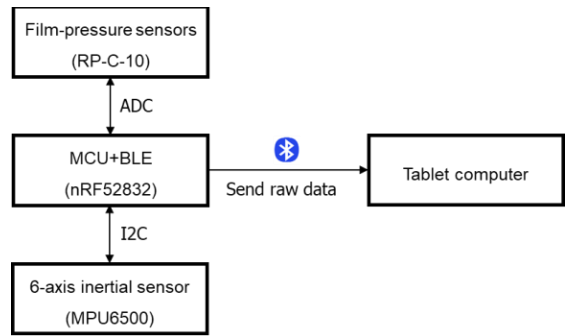

Fig. 6. Working schematic of the system

TABLE I. SPECIFICATION LIST OF THE PROTOTYPE DEVICE

| Name | Spec |
|---|---|
| inertial sensor | MPU6500, 6-axis, InvenSense |
| film-pressure sensor | RP-C-10, <5KG, Film-sensor |
| battery | 3.7V, 300mAH, Li-ion |
| Charing ic | TP4057, TOP POWER ASIC |
| MCU+BLE | nRF52832, Nordic |

### III. THE GAIT PARAMETER ESTIMATION ALGORITHM

Based on the proposed system, in this section the gait spatial and temporal parameters estimation algorithm is presented. This algorithm is just for the handling of collected gait raw data by the prototype device, and for actual using of the system, the algorithm was implemented by java programing language as an Google Android application which can install into the tablet computer with Google operation system. Therefore, the algorithm input data are obtained from the presented prototype device : 6-axis inertial sensor's data and film-pressure sensors data. The algorithm are fully using the fusion of these two kinds of sensors signals. And the algorithm outputs are the gait spatial and temporal parameters, which include stance phase, swing phase, double stance phase, cadence, step time, stride time, stride length, velocity.

In this section, we firstly give the definition of the output gait spatial and temporal parameters. Secondly, gait spatial and temporal parameters estimation algorithm by fusion of film-pressure sensor's signal and internal sensor's signal is presented, which include the detection method of zero velocity phase----the key for accurate gait spatial and temporal parameters estimation.

#### A. Gait Parameters Definition

In this paper, the general gait related spatial and temporal parameters, which are important for GA and assessing the regularity, symmetry and mobility of gait[26], are defined.

Before giving the definition of gait parameters, gait cycle(GC) should be explained at first. Gait cycle is the time period during walking in which one foot heel contacts the ground to when that same foot heel again contacts the ground. So gait cycle is divided into left foot's cycle and right foot's cycle. Each foot's gait cycle consists two phase, namely the stance phase and the swing phase. And the stance phase can be further divided into load phase, foot-flat phase(zero velocity phase) and push phase [27]. The diagram of one normal gait cycle is shown in Fig.7.

**Stance phase(%):** it is the time period during walking in which one foot heel contacts the ground to when that the same foot toe is off the ground. So stance phase is divided into left foot's and right foot's($STP_{left}, STP_{right}$). In this paper, the proportion of stance phase in whole gait cycle is defined as this formula:

$$STP = \frac{toeoff - heelstrike}{GC} \times 100\% \quad (1)$$

**Swing phase(%):** it is the time period during walking in which one foot toe is off the ground to when that the same foot heel contacts the ground. So swing phase is divided into left foot's and right foot's($SWP_{left}, SWP_{right}$). In this paper, the proportion of swing phase in whole gait cycle is defined as this formula:

$$SWP = \frac{heelstrike - toeoff}{GC} \times 100\% \quad (2)$$

**Double stance phase(%):** it is the time period when both feet are in contact with the ground during one gait cycle. In this paper, the proportion of double stance phase in whole gait cycle is defined as this formula:

$$DSP = \frac{time\{both\ feet\ in\ stane\ phase\}}{GC} \times 100\% \quad (3)$$

**Cadence (steps/min):** it is step number per minute. Let N be the number of steps taken over the time period t (in second). Cadence can thus be expressed:

$$CAD = 60\ \frac{N}{t} \quad (4)$$

Normally human's cadence is 95~125 steps/min.

**Step time(s):** it is the time period during walking in which one foot heel contacts the ground to when that the other foot heel contacts the ground. So step time is divided into left foot's and right foot's ($STT_{left}, STT_{right}$).

$$STT_{right} = heelstrike_{left} - heelstrike_{right} \quad (5)$$
$$STT_{left} = heelstrike_{right} - heelstrike_{left} \quad (6)$$

In normal gait, the step time of left foot is equal to the step time of right foot.

**Stride time(s):** it is the time period during walking in which one foot heel contacts the ground to when that the same foot heel contacts the ground again. It is equal to gait cycle. Hence, the stride time is the sum of $STT_{left}$ and $STT_{right}$:

$$STRT = STT_{left} + STT_{right} \quad (7)$$

**Stride length(cm):** it is defined as the distance between any two successive points of heel contact of the same foot. In a normal gait, the stride length is the walking distance during one whole gait cycle.

**Velocity(km/h):** Gait velocity has been consistently shown to be an important indicator and predictor of health status, especially in older adults [28,29]. It is the speed of walking. Let $L$ be the distance(in centimeter) of walking taken over the time period t (in second). Gait velocity can be expressed:

$$VGait = \frac{3.6}{100} \frac{L}{t} \quad (8)$$

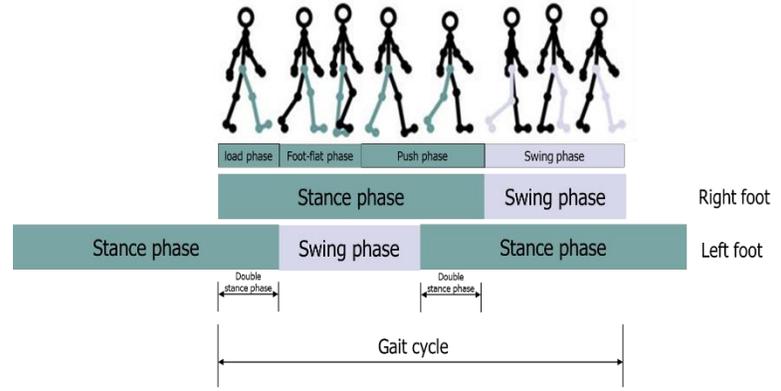

Fig. 7. Diagram of one gait cycle

Regarding to the definition of these gait parameters, in this paper these parameters are classified as follows: Stance phase, swing phase, double stance phase, cadence, step time and stride time belong to gait temporal parameters; Stride length and velocity belong to gait spatial parameters.

*B. Algorithm for Gait Spatial and Temporal Parameters*

As we know, the basic and key process for gait analysis is to detect each footstep's heel event(heel strike & heel off) and toe event(toe on & toe off). After detecting these events, more gait temporal parameters can be calculated accordingly. In addition, accurate position of foot while walking are calculated by double integrating acceleration data, which has been corrected for drift problem by using zero velocity update. Hence the detection of zero velocity phase is very important for accurate gait spatial parameters. Therefore, to detect heel event and toe event is the key for gait spatial and temporal parameters(In fact, zero velocity is calculated by the time position of toe on and heel off : $heeloff - toeon$). In this section, a fusion algorithm by using film-pressure sensor's signal and internal sensor's signal is presented to detect heel event and toe event. The diagram of the algorithm is shown in Fig. 8.

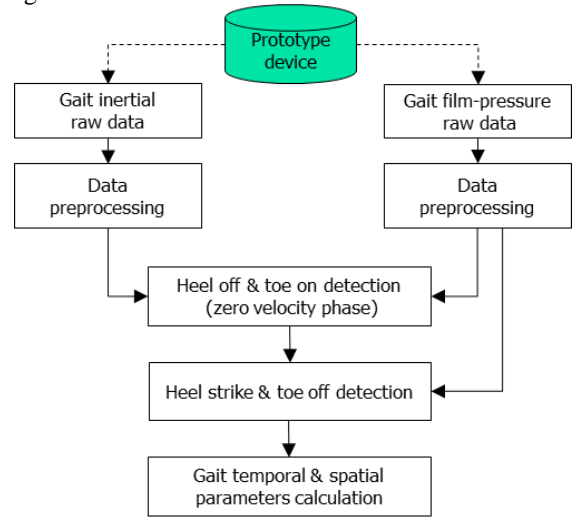

Fig. 8. Diagram of gait parameters estation algorithm

*1) Preprocessing of data*

Before handling of the raw data from two kinds of sensors, data preprocessing is needed. At first, for inertial sensor signal handling, in this work we use body coordinates shown in Fig. 9 (y-axis is toward the walking direction).

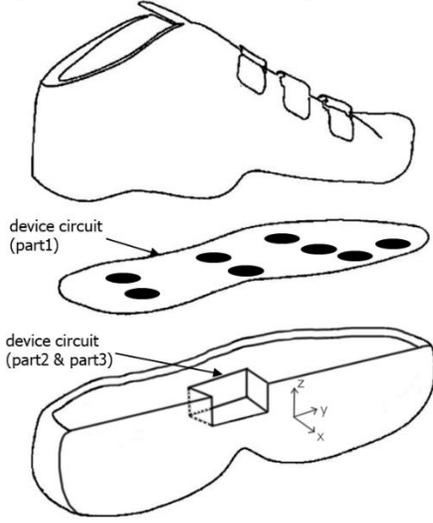

Fig. 9. Diagram of coordinates & assembling of main circuits(left foot as example)

In addtion, regarding to the 6-axis inertial sensor signals, in order to smooth the inertial sensor's signal, a low-pass-filter (LPF) is provided by chipset provider and is designed as follows according to the datasheet's specification of the chipset:

$$y(n) = \sum_{i=0}^{N}(w_i x(n)) \qquad (9)$$

Where $x(n)$ is the input signal, N is the number of coefficients, and $w_i$ is the coefficients of LPF. Specifically, in this work, cut frequency is 20 Hz.

Regarding to the film-pressure sensor's signals, here a Gauss filter (GF) is designed ($\delta$ is 5, number of coefficients is 7) for smooth the film-pressure sensor's signal:

$$y(n) = \frac{1}{\sqrt{2\pi}\delta} e^{-\frac{x(n)^2}{2\delta^2}} \qquad (10)$$

Where $x(n)$ is the input signal.

*2) Detection of zero velocity*

In fact, zero velocity is calculated by the time position of toe on and heel off : $heeloff - toeon$. Hence in this part, we mainly focus on how to detect the event of heeloff and toeon.

Here we define that channel 2, 3, 4 of film-pressure sensors is the pressure value of forefoot, and channel 7, 8 of film-pressure sensors is the pressure value of hind foot. Set $P_1$ is the average pressure value of forefoot, and $P_2$ is the average pressure value of hind foot, so the pressure sum of whole foot(forefoot and hind foot) is :

$$P = P_1 + P_2 \qquad (11)$$
$$P_1 = \frac{1}{3}\sum channel\ 2,3,4 \quad, P_2 = \frac{1}{2}\sum channel\ 7,8 \qquad (12)$$

In order to get more smooth and obvious signal of film-pressure sum signal($P$), we let it do a LPF again as the formula(9) showed. Specifically, here's cut frequency is 0.02 Hz, and number of coefficients is 23.

As we know, the start point of zero velocity phase(toeon) happens during the increasing of pressure sum($P$), and the end point of zero velocity phase(heeloff) happens during the decreasing of pressure sum($P$). Thus, to get out these two points, we can combine the inertial sensor's signal,especially the X-axis of gyroscope's signal. As we all know that signal's changing degree (signal's fluctuations) can be easily indicated by variance:

$$s^2 = \frac{\sum_{i=1}^{n}(x_i - x)^2}{n}, x = \frac{\sum_{i=1}^{n}x_i}{n}, x_i = a_i - a_{i-1} \qquad (13)$$

$$\begin{cases} s^2 \geq variance_{threshold1}, & \text{it is heel off event} \\ s^2 < variance_{threshold2}, & \text{it is toe on event} \end{cases} \qquad (14)$$

where $a_i$ is the gyroscope's X-axis signal, n is the sample number gyroscope's X-axis signal. Therefore, the start point of zero velocity phase(toe on) is that the change of gyroscope's X-axis signal is very small during the increasing of pressure sum($P$), namely, $s^2$ is very small; The end point of zero velocity phase(heel off) is that the change of gyroscope's X-axis signal is very big during the decreasing of pressure sum($P$), namely, $s^2$ is very big. After using the signals of film-pressure( $P$) and signals of inertial sensor(gyroscope's X-axis signal), the zero velocity phase can be detected as it is shown in Fig. 10.

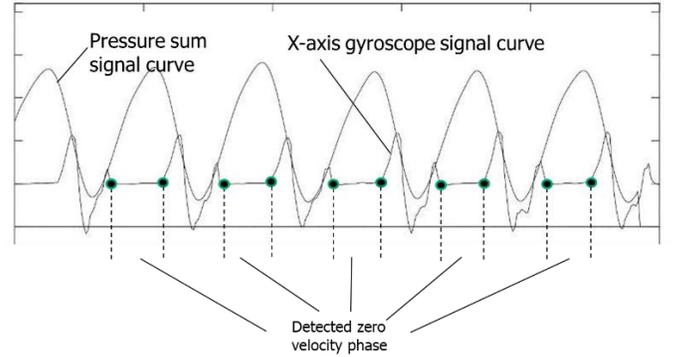

Fig. 10. Diagram of zero velocity phase detection

*3) Detection of heel strike & toe off*

Regarding to detecting the event of heel strike and toe off, we will use and refer to the result of zero velocity phase. As we know, heel strike event usually happens a little early than the start point of zero velocity phase, and toe off event usually happens a little late than the end point of zero velocity phase. For heel strike event detection, we will use the $P_2$ signal because it represents the pressure of hind foot, and for toe off event detection, we will use the $P_1$ signal because it represents the pressure of forefoot. As it's shown in Fig. 11, green color signal curve is $P_1$, blue color signal curve is $P_2$, and red color signal curve is the detected zero velocity phase. As we know, heel strike event has the characteristic that the pressure signal of hind foot ($P_2$) is from flat trend to steep increasing trend, and toe off event has the characteristic that the pressure signal of forefoot ($P_1$) is from steep decreasing trend to flat trend. The specific algorithm is as follows:

Select 66 samples from $P_2$ ($P_1$) and these samples are near the start(end) point of zero velocity phase: 60 samples before(after) the time of start(end) point of zero velocity phase, and 6 samples after(before) the time of start(end) point of zero velocity phase. Each three samples (we also call it 'point' in signal curve) can create an angle θ(Fig. 12), and generally, angle θ have four kinds of possible shape (Fig. 13). Then we can set ΔC:

$$\Delta C = \begin{cases} 1 - \cos(|\theta - 180°|), \theta < 180° \\ \cos(|\theta - 180°|) - 1, \theta > 180° \end{cases} \quad (15)$$

Hence the bigger ΔC is, the bigger the changing of pressure signal is. Then we select the biggest three ΔC ($\Delta C^1, \Delta C^2, \Delta C^3$)'s corresponding angle θ ($\theta^1, \theta^2, \theta^3$)'s middle point as the candidates heel strike (toe off) event point. Therefore, in the followings, we need to find out which angle θ ($\theta^1, \theta^2, \theta^3$) or which angle's middle point is our real heel strike (toe off) event point. In our work, we use three score methods to evaluate these candidates from three different modes (set full score be 100):

**SCORE 1($m$):** it is from the mode of horizontal distance between candidate points and start(end) point of zero velocity phase. We set coordinate in Fig. 11, and coordinate 0 point is at the left-down of the figure. Let $X_1, X_2, X_3$ is the horizontal distance of angle θ ($\theta^1, \theta^2, \theta^3$)'s middle point to the start(end) point of zero velocity phase, then the score of these three candidates are:

$$\begin{cases} m_1 = -\left(\frac{X_1}{\max(X_1, X_2, X_3)} \times 100\right) \\ m_2 = -\left(\frac{X_2}{\max(X_1, X_2, X_3)} \times 100\right) \\ m_3 = -\left(\frac{X_3}{\max(X_1, X_2, X_3)} \times 100\right) \end{cases} \quad (16)$$

The bigger this score is, the more near the candidate point to the start(end) point of zero velocity phase.

**SCORE 2($n$):** it is from the mode of changing degree of left points' signals of the candidate point. The bigger this score, the bigger the signal's changing degree, namely, the more steep trend of the pressure signal's curve. Let $r$ is the number of selected points of the candidate point's left, and then the selected points is $C_1, C_2, C_3, ..., C_r$. Let $b_{X_i}$ is the vertical coordinates of the candidate point($X_i$), then the vertical coordinates of the selected points are $b_1, b_2, b_3, ..., b_r$. Set,

$$S(X_i) = \sum_{B=b_1}^{b_r} |B - b_{X_i}| \quad (17)$$

So the score of these three candidates are:

$$\begin{cases} n_1 = \left(\frac{S(X_1)}{\max(S(X_1), S(X_2), S(X_3))} \times 100\right) \\ n_2 = \left(\frac{S(X_2)}{\max(S(X_1), S(X_2), S(X_3))} \times 100\right) \\ n_3 = \left(\frac{S(X_3)}{\max(S(X_1), S(X_2), S(X_3))} \times 100\right) \end{cases} \quad (18)$$

**SCORE 3($l$):** it is from the mode of changing degree of right points' signals of the candidate point. The bigger this score, the bigger the signal's changing degree, namely, the more steep trend of the pressure signal's curve. Let $r$ is the number of selected points of the candidate point's right, and then the selected points is $C_1, C_2, C_3, ..., C_r$. Let $b_{X_i}$ is the vertical coordinates of the candidate point($X_i$), then the vertical coordinates of the selected points are $b_1, b_2, b_3, ..., b_r$. Set,

$$S(X_i) = \sum_{B=b_1}^{b_r} |B - b_{X_i}| \quad (19)$$

So the score of these three candidates are:

$$\begin{cases} l_1 = \left(\frac{S(X_1)}{\max(S(X_1), S(X_2), S(X_3))} \times 100\right) \\ l_2 = \left(\frac{S(X_2)}{\max(S(X_1), S(X_2), S(X_3))} \times 100\right) \\ l_3 = \left(\frac{S(X_3)}{\max(S(X_1), S(X_2), S(X_3))} \times 100\right) \end{cases} \quad (20)$$

In general, for detecting of heel strike event point and toe off event point from the three candidates, we use the total score that the biggest score among the candidate points, and the corresponding point is just the heel strike event point or the toe off event point.

$$score_{heelstrike} = \max_{i=1,2,3}\{m_i - n_i + l_i\} \quad (21)$$

$$score_{toeoff} = \max_{i=1,2,3}\{m_i + n_i - l_i\} \quad (22)$$

According to this method, the detected event of heel strike is blue circle point, and the detected event of toe off is the green circle point(Fig. 11).

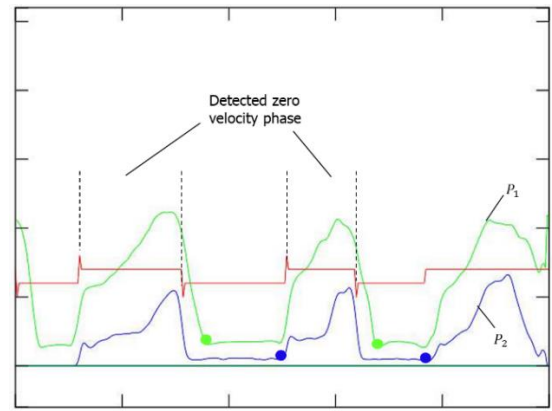

Fig. 11. Diagram of pressure signal of forefoot and hind foot

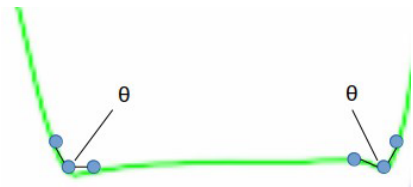

Fig. 12. Diagram of angle by each three pressure samples

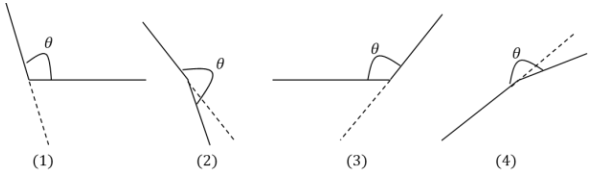

Fig. 13. Diagram of possible angle(1&2 is toe off event, 3&4 is heel strike event)

## IV. EXPERIMENTS

To show the effectiveness of the proposed system and algorithm of the paper, an experiment is performed with 27 subjects from local hospital(China Rehabilitation Research Center). These subjects were diagnosed as stroke illness by the hospital, and 17 of them are hemiplegia with left side, and 10 of them are hemiplegia with right side. Before the experiment, all the subjects were told the purpose, process, and possible harm of the experiment. Each participating subject selects the available size of the prototype device(at present, we have developed three kinds of size : small, middle, big) according to his/her foot size, and walks straight in the 3X0.6m rectangle test ground two times. And the time interval of each subject's measurement by our proposed system and by the compared GA tool is very small(normally less than 2 minutes). For the details of these attended subjects, please refer to Table II.

TABLE II.　　DETAILS OF EXPERIMENT ATTENTED SUBJECTS

| Number | male | female | Ages | Weight (kg) | Height (cm) |
|---|---|---|---|---|---|
| 27 | 16 | 11 | 50 | 68.80 | 166.70 |

In the experiment, the spatial-temporal gait parameters obtained with the system (Gaitboter) and the algorithm of this paper are compared with a GA tool (MyoMotion MR3 3.8.6, Noraxon U.S.A., Inc.) using at department of rehabilitation assessment in China Rehabilitation Research Center. And the experimental results are shown in Table III.

TABLE III.　　CAMPARING RESULT OF GAIT PARAMETERS

| Parameters | Gaitboter | MyoMotion | |diff| |
|---|---|---|---|
| Stancephase,%,Left | 78.741±9.047 | 80.315±8.324 | 1.574 |
| Swing phase,%,Left | 21.259±9.047 | 19.685±8.324 | 1.574 |
| Stance phase,%,Right | 82.819±8.117 | 82.552±7.422 | 0.267 |
| Swing phase,%,Right | 17.181±8.117 | 17.448±7.422 | 0.267 |
| Double stance,% | 61.607±8.640 | 64.233±11.730 | 2.626 |
| Step time,sec,Left | 1.157±0.458 | 1.252±0.639 | 0.095 |
| Step time,sec,Right | 1.034±0.355 | 1.216±0.563 | 0.182 |
| Stride time,sec | 2.176±0.762 | 2.466±1.133 | 0.290 |
| Cadence,step/min | 62.150±19.834 | 57.370±20.681 | 4.780 |
| Stride length,cm | 48.544±14.946 | 40.815±13.01 | 7.730 |
| Velocity,km/h | 0.910±0.48 | 0.711±0.423 | 0.199 |

From the table III, we can see that all the parameters measured by both systems(Gaitboter & MyoMotion) are the mean value ± standard deviation(measurement error) of all participating subjects. Almost all the parameters estimated by the proposed system(Gaitboter) is consistent with the characteristics reported in the paper [30] that for stroke patients' gait there is slow speed of walking, long stance phase, etc. At the same time, we also gave the absolute value of the deviation(|**diff**| at the table III) between the measured mean value of the proposed system(Gaitboter) and the measured mean value of the GA tool(MyoMotion). From the results of |**diff**|, the |**diff**| of each parameter is smaller than its corresponding standard deviation no matter the proposed system(Gaitboter)'s standard deviation or the GA tool(MyoMotion)'s standard deviation. This result gives the fact that the measurements of the proposed system(Gaitboter) comparing with the GA tool(MyoMotion)'s measurements is considered acceptable. In addition, as we know that Pearson correlation coefficient (PCC) is a measure of the linear correlation between two variables. Therefore, we calculated the PCC of these two set of gait parameters measured by Gaitboter and MyoMotion respectively that PCC is 0.996014059. This result strongly proves the consistency and correlation of these two gait analysis tools.

## V. CONCLUSION AND FUTURE WORK

In this article, we present a two-foot-worn in-shoe system (Gaitboter) based on low-cost, wearable and multimodal sensors' fusion for GA, comprising an inertial sensor and eight film-pressure sensors with each foot for gait raw data collection while walking. And then based on the system, a GA algorithm for estimating the spatial-temporal parameters of gait is proposed. The algorithm fully uses the fusion of two kinds of sensors' signals: inertial sensor and film-pressure sensor, in order to estimate the spatial-temporal gait parameters, such as stance phase, swing phase, double stance phase, cadence, step time, stride time, stride length, velocity. Finally, to verify the effectiveness of the system and algorithm of the paper, an experiment is performed with 27 stoke patients from local hospital that the spatial-temporal gait parameters obtained with the system and the algorithm are compared with a GA tool used in medical laboratories. The experimental results show that it achieves very good consistency and correlation between the proposed system and the compared GA tool, moreover, the measured results by the proposed system is considered acceptable. It proves the proposed system and the fusion algorithm of two sensor's signal is beneficial and effective for wearable GA.

In future work, for generalization of our study we will add more kinds of subjects such as multiple sclerosis, Parkinson's disease, etc. And we will consider to implement the plantar pressure analysis with the system's all film-pressure points(channels).

ACKNOWLEDGMENT

This work is supported by Beijing Natural Science Foundation (4172058) and partly supported by National Key Technology R&D Program of China (2014BAK15B02).